# Coevolution of broad emission lines and X-ray spectrum in changing-look AGNs

Hao Liu,[1,2] Qingwen Wu,[1] and Bing Lyu[1,3]

[1]*School of Physics, Huazhong University of Science and Technology, Wuhan 430074, People's Republic of China*
[2]*School of Astronomy and Space Science, University of Science and Technology of China, Hefei 230026, People's Republic of China*
[3]*Shanghai Astronomical Observatory, Chinese Academy of Sciences, Shanghai, 200030, People's Republic of China*

## ABSTRACT

Changing-look active galactic nuclei (CLAGNs) show the disappearance and reappearance of broad emission lines in a few years, which challenges the orientation-based AGN unification model. We reduce the X-ray data for five well-studied CLAGNs that show a strong change in broad emission lines in the past several decades. We find that the X-ray photon index, Γ, and the Eddington-scaled X-ray luminosity, $L_{2-10\text{keV}}/L_{\text{Edd}}$, normally follow negative and positive correlations when the Eddington ratio is lower and higher than a critical value of $\sim 10^{-3}$. We find that the CLAGNs observed with broad H$\beta$ emission lines stay in the positive part of the Γ–$L_{2-10\text{keV}}/L_{\text{Edd}}$ correlation, while the broad H$\beta$ lines become weak or disappear in the anticorrelation part of the Γ – $L_{2-10\text{keV}}/L_{\text{Edd}}$ correlation, which suggests that the evolution of the broad lines should be correlated with the evolution of the underlying accretion process. We further find that the CLAGNs are consistent with the other different types of AGNs in the $L_{\text{bol}}$–$L_{\text{bol}}/L_{\text{Edd}}$ correlation. These results support that the CLAGNs are belong to a special stage of AGNs with a bolometric Eddington ratio ~1%, where the broad emission lines are easily affected by the strong variation in ionization luminosity that is caused by the transition of accretion modes.

*Keywords:* Active galactic nuclei(16) — Accretion(14) — Black hole physics(159)

## 1. INTRODUCTION

An active galactic nucleus (AGN) is defined as a galaxy containing an accreting supermassive black hole (SMBH) with an Eddington-scaled bolometric luminosity $L_{\text{bol}}/L_{\text{Edd}} > 10^{-5}$(Yuan & Narayan 2014, ; $L_{\text{Edd}}$ is the Eddington luminosity), where the nuclear activity is much different from that of normal galaxies. A fraction of AGNs emits strong and broad emission lines (BELs) with an FWHM from ~1000 to ~10,000 km $s^{-1}$ (Peterson 1997), while other AGNs are observed with only narrow emission lines (NELs) with an FWHM lower than ~1,000 km $s^{-1}$. These two types of AGNs are classified as type 1 AGNs and type 2 AGNs, respectively. Both BELs and NELs are mainly photoionized by the accretion disk, where the line ratios support a nonstellar origin (e.g., Netzer 2015). There is a putative clumpy torus outside of the subparsec broad-line region (BLR) and within the kiloparsec narrow line region, and it is widely believed that the type 1 AGNs are viewed face-on to the opening of the torus, while in type 2 AGNs, one faces the obscuring part (e.g., Antonucci 1993; Urry & Padovani 1995). The dust grains in the torus absorb optical/ultraviolet (UV) photons from the accretion disk and re-emit in the IR band, which is the dominant source of near- to mid-IR radiation in most AGNs. Some type 2 AGNs show broad lines in their polarized spectra (e.g., NGC 1068; Antonucci & Miller 1985), which further supports this orientation-based unification model. It should be noted that not all type 2 AGNs show the broad lines in their polarized spectra (Tran 2001), which suggests that the AGN unification depends not on the viewing angle alone. Elitzur & Ho (2009) proposed that the broad lines in different types of AGNs follow an evolutionary sequence of type1.0→1.2/1.5→1.8/1.9→2.0 as the accretion rate or Eddington ratio decreases. The intermediate AGN type (e.g., type 1.5-1.8/1.9) shows a weak/absent broad H$\beta$ lines and retains a broad H$\alpha$ component (Osterbrock 1981).

Corresponding author: Qingwen Wu
qwwu@hust.edu.cn

Changing-look AGNs (CLAGNs) are AGNs that show transitions from one type to another within a time-scale of years or decades. With more and more multiband surveys, the number of CLAGNs has grown rapidly in the last few years (e.g., McElroy et al. 2016; Yang et al. 2018; Shapovalova et al. 2019). The term "changing look" was originally referred to X-ray observations of Compton-thick AGN (column density $N_H \geq 10^{24}$ cm$^{-2}$) becoming Compton-thin ($N_H \leq 10^{24}$ cm$^{-2}$), and vice versa (e.g., Ricci et al. 2016, and references therein). In optical observations, some AGNs undergo a dramatic variability of the continuum, emission line profiles, and classification types (e.g., from type 1 to type 2, or vice versa); these are also defined as CLAGNs. The physical mechanism for the CLAGNs is still unclear, which can play a key role in understanding the nature of AGNs. The disappearance/reappearance of BELs or the strong variation in column density are ascribed to the variation in the clumpy torus, where the variable absorber moves in or out from our line of sight (e.g., Goodrich 1989). An alternative scenario is that the changing-look process is triggered by the intrinsic change in the central engines of these AGNs (e.g., Noda & Done 2018; Stern et al. 2018). The varied column density as found in some CLAGNs (e.g., NGC 1365 and IC 751) supports the scenario of a variation in the clumpy torus scenario (e.g., Risaliti et al. 2009; Ricci et al. 2016, and references therein). However, most of the other CLAGNs cannot be explained by the obscuration scenario because their intrinsic absorption is always low (e.g., Mrk 1018; McElroy et al. 2016).

The big blue bump as observed in quasistellar objects (QSOs) and bright Seyferts is always explained by the standard optically thick and geometrically thin accretion disk (Shakura & Sunyaev 1973). The Xray emission is also ubiquitous in AGNs, and is produced by Comptonization of optical/UV disk photons by hot electrons in the corona above the accretion disk. This Comptonization produces a cutoff power-law Xray emission with a typical photon index of $\sim 2$ and a cutoff energy $\sim 100$ keV (see Courvoisier 2013). The optically thick disk can be irradiated by the X-ray continuum and responds by producing an X-ray reflection spectrum, where the most important features are the broad Fe K$\alpha$ lines at 6.4 keV and a hard X-ray hump peaking at 20 – 30 keV as observed in many bright AGNs. It should be noted that the narrow Fe K$\alpha$ lines (FWHM < several thousand km $s^{-1}$) are also common in AGNs, which may originate from the outer part of cold disk, the BLR, and/or torus (e.g., Yaqoob & Padmanabhan 2004). The standard thin disk will transit to the optically thin, geometrical thick advection dominated accretion flow (ADAF) in low-accretion rate regime (e.g., the accretion rate is lower than a few percent of the Eddington accretion rate), which exists in low-luminosity AGNs(e.g., Yuan & Narayan 2014, and references therein). The plasma in ADAF is hot (e.g., $T_e \sim 100$keV), and the thermal Comptonization dominates the hard X-ray emission, where the seed photons for inverse Compton mainly come from the synchrotron radiation. The negative and positive correlations between the X-ray photon index and the Eddington-scaled luminosity as found in low-luminosity AGNs and luminous AGNs could be a signature of this transition from the thin to the thick disk. A similar behavior has been found in X-ray binaries (XRBs; Wang et al. 2004; Wu & Gu 2008; Gu & Cao 2009; Liu et al. 2019). The negative and positive correlations are roughly consistent with the ADAF and disk/corona model, respectively. In ADAF, the optical depth will increase as the accretion rate increases, which will lead to a harder X-ray spectrum (i.e., an anticorrelation) because that the electron temperature is normally around 100 keV (Qiao & Liu 2013, e.g.,). In the high-luminosity regime, the optical depth of the corona will decrease as the accretion rate increases due to the stronger cooling, which will lead to a softer X-ray spectrum (i.e., positive correlation; Cao 2009; You et al. 2012; Qiao & Liu 2013).

The origin, geometry, and kinematics of BELs in AGNs are still unclear. The CLAGNs provide a new window for studying the physics of BELs, where the reappearance and disappearance of BELs can shed light on its physical mechanism. In this work, we explore the possible coevolution of BELs and X-ray spectra based on a sample of five CLAGNs with quasisimultaneous multiple X-ray and optical observations. In Sect. 2, we present the sample and data reduction, the results are presented in Sect. 3, and the discussion and summary are presented in Sect. 4. We adopt a cosmology model of $\Omega_m = 0.32$, $\Omega_\Lambda = 0.68$, and $H_0 = 67$ km s$^{-1}$ Mpc$^{-1}$ (Planck Collaboration et al. 2014) in this work.

## 2. SAMPLE SELECTION AND DATA REDUCTION

Up to now, more than one hundred CLAGNs are reported in the literature, and we have built a large sample of 108 CLAGNs to explore their basic properties (Liu et al. 2022, in preparation). To explore the coevolution of BEL and X-ray spectrum, we choose the CLAGNs with X-ray flux variations stronger than one order of magnitude based on the published data from the literature. We finally find five CLAGNs, which are Mrk 1018 (Lyu et al. 2021), NGC 3516 (Ilić et al. 2020), Mrk 590 (Mathur et al. 2018), NGC 2992 (Marinucci et al.,

**Table 1.** Basic Information of CLAGNs

| Source | Redshift | $\log(M_{\rm BH}/M_\odot)$ | References | $N_{\rm H}/10^{22}$ cm$^{-2}$ | References |
|---|---|---|---|---|---|
| (1) | (2) | (3) | (4) | (5) | (6) |
| Mrk 1018 | 0.0424 | 8.25 | a | 0.02 | b |
| NGC 3516 | 0.0088 | 7.49 | c | 3.00 | d |
| Mrk 590 | 0.0264 | 7.50 | e | 0.03 | b |
| NGC 2992 | 0.0077 | 7.72 | f | 0.85 | g |
| NGC 1566 | 0.0050 | 6.92 | f | 0.01 | b |

Note—Note. Column (1): Source names. Column (2): Redshift. Columns (3) and (4): Black hole masses and references. Columns (5) and (6): Column density of the neutral hydrogen gas $N_{\rm H}$ and references. The references are (a) Winter et al. (2010), (b) Kalberla et al. (2005), (c) Grier et al. (2013), (d) Huerta et al. (2014), (e) calculated from the scaling relations in McConnell & Ma (2013) and from the central velocity dispersion of the source from the web: http://leda.univ-lyon1.fr/, (f) Woo & Urry (2002), and (g) Marinucci et al. (2018)

2018), and NGC 1566 (Oknyansky et al. 2020; Jana et al. 2021). The durations of the type transitions are normally longer than 2-3 yr in these five CLAGNs, and we simply define the quasisimultaneous data if the X-ray and optical observations are within 1 yr of the X-ray ones. Based on the spectra of the broad H$\beta$ line, optical observations with or without evident broad lines are classified as BEL type and NEL type, respectively. Guolo et al. (2021) found that NGC 2992 have BELs in six optical observations from 2014 to 2021, and we define four X-ray observations during this period as BEL type even though the time separation of two observations is slightly longer than one year. The detailed emission line types and their references are listed in Column (4) and (5) of Table 2. We note that some X-ray observations have no quasisimultaneous optical spectral information, and we define these observations as uncertainty type (U), which can still shed light on the X-ray spectral evolution in these CLAGNs. By searching the archive X-ray observations of these five sources from the XMMNewton, Chandra, and NuSTAR, 68 X-ray observations (including 5 simultaneous XMM-Newton/NuSTAR observations) are selected. Among these observations, we find that 19 observations show BEL-type spectra and 31 observations show NEL-type spectra. To compare them, we calculate the absorption-corrected luminosity at 2-10 keV for each observation based on best spectral modeling (see Sect. 2.1-2.3). Except for the Galactic absorption, three CLAGNs show no evident additional absorption component (Mrk 1018, Mrk 590, and NGC 1566), and two other sources need an additional absorption component (NGC 2992 and NGC 3516). In this work, the hydrogen column density $N_{\rm H}$ was fixed as a constant in the spectral fitting, where the value of $N_{\rm H}$ and references are listed in Table 1. The basic parameters (e.g., BH masses and redshift) for CLAGNs are also presented in Table 1. The X-ray data reduction is described in detail as follows, and the fitting results are listed in Table 2.

### 2.1. NuSTAR

There are 24 NuSTAR observations for these five CLAGNs, and the reduction of the NuSTAR is conducted following the standard procedures using the NuSTAR Data Analysis Software (nustardas v2.0.0). We use *nupipeline* and the CALDB 20210315 to produce clean event files. We use the task package *nuproducts* to extract the source and background spectra with most parameters kept as task default. We extract the source counts from a circle of 50 diameter circular region (see also Lyu et al. 2021) centered on the source, and background counts from the same circular region without the source. The 3-78 keV spectra from the FPMA and FPMB instruments are grouped by a minimum of 30 counts per spectral bin. Spectral fitting based on the bin data is done with the software XSPEC (version 12.11.1) in HEASOFT (v6.28) for all observations. To fit the possible reflection component of the Fe K$\alpha$ line and the Compton hump at $\sim$ 20-30 keV, we adopt a reflected power-law component (PEXRAV) with absorption (PHABS), where the gaussian line (ZGAUSS) at 6.4 keV is also added to mimic the possible Fe K$\alpha$ line if it can improve the fitting. For the PEXRAV component, the reflection scaling factor is set as a free parameter, and the cutoff energy is fixed to 200 keV.

## 2.2. *XMM-Newton*

We analyze 30 archival data of five CLAGNs based on EPIC-PN, MOS1, and MOS2 on board XMM-Newton, which are reduced with the Science Analysis Software (SAS, version 19.1.0). The source and background regions are selected within a 10 diameter circular region that includes and excludes the source, respectively. Three observations of NGC 2992 show pile-up effect, and we extract from annular source regions of 10″ and 40″ as inner and outer radius (ID0147920301) or 5″ and 10″ as inner and outer radius (ID0840920301). The spectra of ID0840920201 are extracted from an annular source region of 10″ and 40″ as inner and outer radius for PN and MOS1 instruments, and the spectra observed by MOS2 are extracted in an annular region with 5″ and 10″ as inner and outer radius. The spectra are reduced and fitted in 1-10 keV range for four sources, while the spectra of NGC 3516 are fitted in 3-10 keV due to their complex obscuration in the soft X-ray band below 3 keV (Noda et al. 2016). In the spectral fitting, the reflection component is neglected because it cannot be well constrained based on the narrow energy band of 1-10 keV, so we replace the PEXRAV with a power-law component (POW). We note that 5 of 30 observations have simultaneous NuSTAR observations. The combined spectra are analyzed for these five observations.

## 2.3. *Chandra*

We extract the Chandra ACIS-S spectra with CIAO (version 4.11) and CALDB (version 4.8.4.1). Fourteen observations for Mrk 1018 and Mrk 590 are considered. We neglect the observations of NGC 3516 due to the complex absorption in the soft X-ray bands (e.g., ≤3 keV) and NGC 2992/NGC 1566 due to pile-up effect. One observation of Mrk 1018 (ID12868) also shows a pile-up effect, and an annular source region within 0.5″ and 3″ as inner and outer radius are adopted to extract spectrum. The other observations are extracted from a 3″ radius circle, and the background spectra are extracted from an annulus with 5″ inner radius and 15″ outer radius. The spectral fitting for these observations is similar to that of XMM-Newton as described above, but in the 1-8 keV range.

## 3. RESULTS

We present the $\Gamma$-$L_{2-10\text{keV}}/L_{\text{Edd}}$ relation for each CLAGN in Figure 1, and then compare the CLAGN with other types of AGNs in the $L_{\text{bol}}$-$L_{\text{bol}}/L_{\text{Edd}}$ correlation in Figure 2. The spectral fittings of each CLAGN are presented in more detail as below, where the spectra are presented in Appendix Figures 3-8.

### 3.1. *The X-Ray Spectral Evolution in CLAGNs*

*Mrk 1018*: The X-ray spectral fittings are shown in Figure 3 & 4. We find that a power-law model or a reflected power-law model can fit the XMM-Newton/Chandra and NuSTAR observations, respectively. The narrow Fe K$\alpha$ line is weak in most cases, and the fittings are not improved substantially even when this component is included. The relation between X-ray photon index, $\Gamma$, and Eddington-scaled X-ray luminosity, $L_{2-10\text{keV}}/L_{\text{Edd}}$, is presented in the top left panel of Figure 1. Three observations before 2010 are observed with BELs (solid triangles), which show a positive correlation of $\Gamma$ = (0.69 ± 0.09) log ($L_{2-10\text{keV}}/L_{\text{Edd}}$) + (3.48 ± 0.22) with Spearman rank correlation coefficients $r = 0.99$ ($p < 10^{-6}$). After 2016, the source entered into the faint state and the BELs are disappear. We find that $\Gamma$ and $L_{2-10\text{keV}}/L_{\text{Edd}}$ follow a negative correlation, where $\Gamma$ = (−0.09 ± 0.10) log($L_{2-10\text{keV}}/L_{\text{Edd}}$) + (1.40 ± 0.32) with $r = -0.46$ ($p = 0.07$). The negative correlation is not very strong, which may be caused by the quite narrow bin of $L_{2-10\text{keV}}/L_{\text{Edd}}$.

*NGC 3516*: The Seyfert 1 galaxy NGC 3516 also underwent strong spectral changes in recent years. We find that an additional absorption component is needed besides the Galactic absorption. We adopt an absorbed power-law model or an absorbed reflected power-law model in this work, where $N_{\text{H}}$ is fixed in our fitting because that there is no strong evidence for variable absorption during variability (e.g., Huerta et al. 2014; Mehdipour et al. 2022). The narrow Fe K$\alpha$ line components are evident in all observations (see Figure 5). Before 2007, the source is observed with BELs. Then, the source transited to the faint state with only NELs during 2014-2019, and returned to the bright state in 2020 and BELs is appear again (Oknyansky et al. 2021). As shown in the top right panel of Figure 1, seven observations with BELs follow a tight positive correlation of the X-ray photon index and Eddington-scaled X-ray luminosity of $\Gamma$ = (1.21 ± 0.08) log($L_{2-10\text{keV}}/L_{\text{Edd}}$) + (5.22 ± 0.21) ($r = 0.96$ with $p < 10^{-3}$). The eight observations with NELs evidently deviate from the above positive correlation, even though they also follow a weak positive correlation of ($\Gamma$ = (0.21±0.12)log($L_{2-10\text{keV}}/L_{\text{Edd}}$)+(2.51±0.42), $r = 0.38$ with $p = 0.35$). The weak correlation for observations with NELs in the faint state may be caused by the large uncertainty due to fewer photons, and more observations are needed to further test this issue.

*Mrk 590*: Three X-ray observations before 2005 are classified as BEL type with $L_{2-10\text{keV}}/L_{\text{Edd}} > 10^{-3}$. Around 9

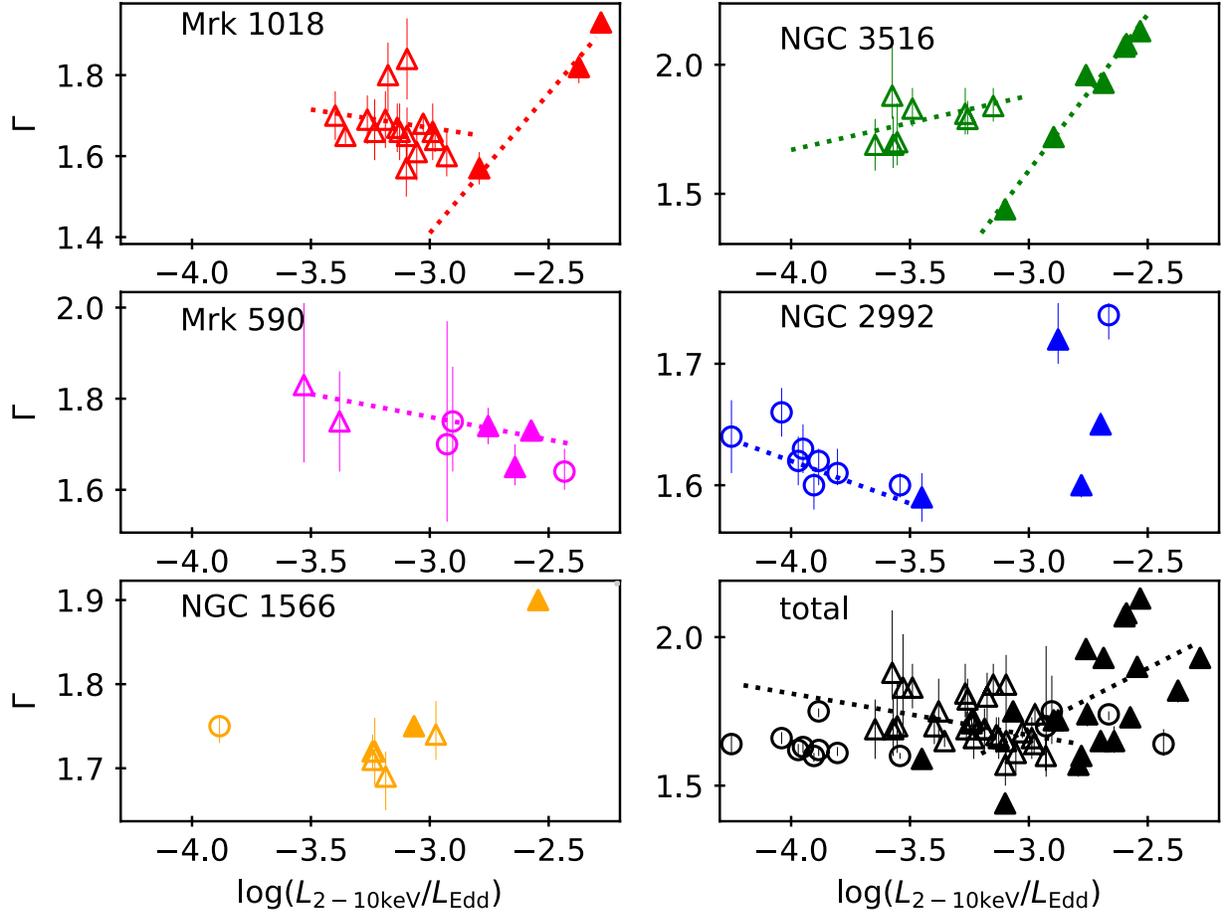

**Figure 1.** The $\Gamma$-$L_{2-10\text{keV}}/L_{\text{Edd}}$ relation for Mrk 1018, NGC 3516, Mrk 590, NGC 2992, NGC 1566, and all sources together. The solid triangles, open triangles, and open circles represent the CLAGNs observed with BELs, without BELs, and with the uncertainty type due to no quasisimultaneous optical spectrum observations, respectively.

years later, it entered into a faint state with only NELs, and it brightened again with $L_{2-10\text{keV}}/L_{\text{Edd}} > 10^{-3}$ after 2016. The X-ray spectral fittings are shown in Figure 6. We adopt an absorbed power-law model for XMM-Newton/Chandra observations or an absorbed reflected power-law model for NuSTAR data, where the absorption is fixed at the Galactic value. The Fe K$\alpha$ line components are added in four observations. We present the $\Gamma$-$L_{2-10\text{keV}}/L_{\text{Edd}}$ of Mrk 590 in middle left panel of Figure 1, and all observations seem to follow a negative correlation ($\Gamma = (-0.10 \pm 0.06) \log(L_{2-10\text{keV}}/L_{\text{Edd}}) + (1.46 \pm 0.15)$, with $r = -0.79$, $p = 0.02$). It should be noted that there are only two observations with NELs and three observations with BELs, which prevents us from further exploring the possibly positive or negative correlations as found in other sources.

*NGC 2992*: The X-ray spectrum of NGC 2992 can be well fitted by an absorbed power-law model, where the fittings are presented in Figure 7. We find that an additional moderate absorption component is needed besides the Galactic absorption. In all 13 observations, the spectrum also presents a narrow Fe K$\alpha$ component at 6.4 keV, which becomes more evident in high-flux cases. There are 4 X-ray observations with quasisimultaneous BELs, but the other observations have no optical spectral information, and we define these observations as uncertain type. The relation between $\Gamma$ and $L_{2-10\text{keV}}/L_{\text{Edd}}$ is shown in the middle right panel of Figure 1. There is no evident correlation of the 4 observations with BELs due to the large scatter, where all these observations have $L_{2-10\text{keV}}/L_{\text{Edd}} > 10^{-3.5}$. Eight uncertain-type observations with $L_{2-10\text{keV}}/L_{\text{Edd}} < 10^{-3.5}$ follow a negative correlation of

($\Gamma$ = (−0.07±0.03) log($L_{2-10keV}/L_{Edd}$)+ (1.34 ± 0.10), $r$ = −0.81 with $p < 0.01$).

*NGC 1566*: NGC 1566 is a famous variable source that shows a faint X-ray emission before 2015. It awaked around 2018 and entered into the faint state again in 2019. We reduce the X-ray data for eight observations from 2015 to 2019, where the absorption is very weak in all observations. The spectra are plotted in Figure 8. There are two observations with observed BELs during 2018, where $L_{2-10keV}/L_{Edd} > 10^{-3.1}$. Another five observations with only NELs have $L_{2-10keV}/L_{Edd} \sim 10^{-3.2}$ (see Table 2). The correlation between $\Gamma$ and $L_{2-10keV}/L_{Edd}$ is shown in the bottom left panel of Figure 1. The two observations with BELs follow a positive correlation, where we do not fit them because there are only two data points. It is the same case as the observations with NELs, where the data points have a similar Eddington-scaled X-ray luminosity and cannot be fit reasonably.

The results for all the observations of five sources are plotted in the bottom right panel of Figure 1. It is evident that BELs have much higher Eddington ratios than NELs. We find that the 19 BEL data points (solid triangles) and 31 NEL data points (open triangles) of five sources follow a positive correlation of ($\Gamma$ = (0.41±0.14) log($L_{2-10keV}/L_{Edd}$) + (2.92±0.38)) and a negative correlation of ($\Gamma$ = (−0.14±0.07) log($L_{2-10keV}/L_{Edd}$) + (1.25±0.21)), where the Spearman rank correlation coefficients are $r$ = 0.66 ($p < 0.01$) and $r$ = −0.46 ($p < 0.01$), respectively. The positive and negative correlations cross at a critical point of log($L_{2-10keV}/L_{Edd}$)∼ −3.02.

### 3.2. *The Eddington Ratio for CLAGNs and Comparison with Other AGNs*

In Figure 2, we compare the CLAGNs with different types of AGNs based on the relation of the bolometric luminosity and the bolometric Eddington ratio ($L_{bol}$ − $L_{bol}/L_{Edd}$). The bolometric luminosity of CLAGNs is derived from the X-ray luminosity in the 2-10 keV band by the empirical relation $L_{bol} = 15.8 L_{2-10keV}$ (e.g. Ho 2009). We find that $L_{bol}/L_{Edd}$ ranges from ∼ $10^{-3}$ to $10^{-1}$ for these CLAGNs, where the CLAGNs with BELs and NELs are roughly divided by $L_{bol}/L_{Edd} \sim 10^{-2}$. The CLAGNs stay in a transition region, where the broad lines evidently become weaker as the Eddington ratio decreases (see the top panel of Figure 2). The BEL CLAGNs are correspond to type 1.0-1.2 AGNs with $L_{bol}/L_{Edd}$ & $10^{-2}$, while NEL CLAGNs are belong to a transition type ∼ 1.5 with $L_{bol}/L_{Edd} \sim 10^{-3}$–$10^{-2}$. The broad lines (e.g., H$\alpha$) may fully disappear for $L_{bol}/L_{Edd}$ . $10^{-5}$, which corresponds to the intrinsic type 2 AGNs.

### 4. SUMMARY AND DISCUSSION

We investigate the possible coevolution of BELs and the X-ray spectra in a sample of CLAGNs with multiple X-ray and optical observations. We find that the CLAGNs with BELs follow a positive $\Gamma$-$L_{2-10keV}/L_{Edd}$ correlation, while CLAGNs with NELs stay in the negative part of the $\Gamma$-$L_{2-10keV}/L_{Edd}$ correlation. Compared with a sample of different types of AGNs, we find that CLAGNs with BELs and NELs follow a similar trend as type 1.0-1.2 and type 1.2-1.9 AGNs, respectively. Our results provide possible evidence that both the disappearance/reappearance of BELs and the evolution of X-ray spectra are triggered by the evolution of the accretion disk.

The CLAGNs in which BELs disappear or appear within several years challenge the classical orientationbased AGN unification model. The physical mechanism that triggered the multiple AGN types as found in the same source at different epochs is unclear. Variable absorption may play a role in some CLAGNs, especially in the X-ray CLAGNs with an observed variable $N_H$ (e.g. Risaliti et al. 2009; Ricci et al. 2016). However, some CLAGNs show intrinsic weak absorption during the changing-look stage (e.g., Mrk 1018, NGC 1566, Noda & Done 2018; Parker et al. 2019). LaMassa et al. (2015) found that the crossing timescale for the obscuration cloud is longer than 20 yr for $M_{BH} = 10^8 M_\odot$ when the intervening object is assumed to stay beyond the sublimation radius. We find that the crossing timescale is much longer than the timescale of the type transition in four CLAGNs (Mrk 1018, Mrk 590, NGC 3516, and NGC 2992) based on their BH masses. NGC 1566 has a shorter crossing timescale of 6-7 yr based on its lower BH mass $M_{BH} \sim 10^7 M_\odot$. However, the BELs are only appear in the flare state during 2017-2019, which is also inconsistent with the obscuration scenario. The strong variation in absorption-corrected X-ray luminosity and mid-IR luminosity implies that the variation is triggered by the intrinsic change in the central ionization source, where the putative torus absorbed the UV/optical emission from the accretion disk and re-emitted in the IR wave band (e.g., Sheng et al. 2017). The radiative efficiency of the ADAF is much lower than that of a standard disk, and therefore, the disk transition will lead to a stronger variation in disk luminosity, where both BELs and IR emission also change after several months to several years.

The different $\Gamma$-$L_{2-10keV}/L_{Edd}$ correlations we found in different stages of CLAGNs support that the evolution of the accretion disk should play an important role in the physical mechanism that causes the changing look. The X-ray spectral evolution is widely explored in stellar-mass BH XRBs, where the spectral index of the power-law component also follows the negative and positive correlation when its flux is lower and higher than a critical value (e.g., Wu & Gu 2008). The similar transition of $\Gamma$-$L_X/L_{Edd}$ between a positive and negative relation is also found in AGNs (Yuan et al. 2007; Constantin et al. 2009; Fender & Belloni 2012; Trichas et al. 2013). Ruan et al. (2019) researched the spectral evolution for six changing-look quasars with the relations of the spectral index between the ultraviolet and X-ray band $\alpha_{OX}$ and the bolometric luminosity. They obtained a similar transition trend as for XRBs, which is consistent with our results, but in the ultraviolet to X-ray band. The negative and positive correlations between $\Gamma$ and $L_{2-10keV}/L_{Edd}$ as found in CLAGNs are quite consistent with those found in XRBs and AGNs, which supports the hypothesis that the accretion mode may be changed during the changing-look stage. It should be noted that the slopes are also different for negative or positive correlations of each CLAGN, which is the same case for XRBs (e.g., Wu & Gu 2008), and the other parameters (e.g., different magnetic field) may contribute this difference. We show all sources in bottom right panel of Figure 1, where the negative and positive correlations for BELs and NELs of five CLAGNs are evident. However, the slope for the fitting of NELs and BELs may be not physical because that each source follows its own trend. It should be noted that the spectral index, Γ, is degenerate with the column density, $N_H$, in the X-ray spectral modeling based on the observations of XMM-Newton/Chandra. The adopted constant column density in our fittings will not affect our conclusions in three CLAGNs (Mrk 1018, Mrk 590, and NGC 1566) because that their absorptions are negligible. In NGC 3516, the complex obscuration in the soft X-ray band below 3 keV is not included in our fittings, where the variable absorption may contribute in the soft X-ray band (Noda et al. 2016). The absorption is negligible in three CLAGNs, and therefore, our assumption of constant $N_H$ should be reasonable. The correlation of $\Gamma$-$L_X/L_{Edd}$ will change if the absorptions are indeed vary in NGC 3516 and NGC 2992. The reflection component of the possible Compton hump in the PEXRAV model is considered in all NuSTAR observations. However, this component cannot be well constrained in XMM-Newton and Chandra observations (e.g., 1-10 keV band). We find that the power-law spectral index may differ by several percent, which will not affect our main conclusion on the $\Gamma$-$L_X/L_{Edd}$ relation.

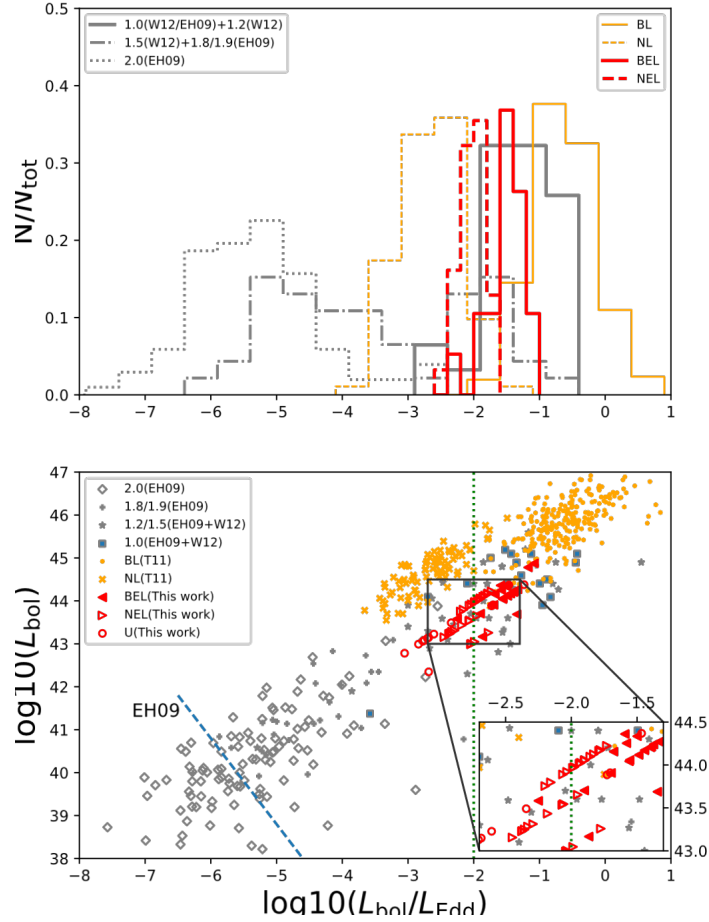

**Figure 2.** $L_{bol}$ and $L_{bol}/L_{Edd}$ relation for CLAGNs and different types of AGNs. The solid and open triangles represent the CLAGNs with evident broad H$\beta$ (BEL type) and no broad H$\beta$ (NEL type) in this work, respectively, where open circles represent U (uncertainty type). For comparison, we also present the results for different types of AGNs (e.g., BL and NL in T11 (Trump et al. 2011), and types 1.0, 1.2/1.5, 1.8/1.9, and 2.0 from EH09 (Elitzur & Ho 2009) and W12 (Winter et al. 2012)). The top panel shows the histogram of each pollution.

The critical Eddington ratios of the negative and positive correlation at $L_{bol}/L_{Edd} \sim 10^{-2}$ are roughly consistent with the model prediction of the accretion mode transition ($\dot{M}/\dot{M}_{Edd} \sim 10^{-2}$; Yuan & Narayan 2014, and references therein). In the case of a hot accretion flow at a low accretion rate, the optical depth of the ADAF for Comptonization decreases as the accretion rate decreases, which will lead to a softer X-ray spectrum or a negative $\Gamma$-$L_{2-10keV}/L_{Edd}$ correlation. As the accretion rate increases, cold clumps and/or even a cold inner disk will be formed in the inner part of the hot accretion flow, and the disk will transit to the cold standard disk. The optical depth of the corona will decrease as more hot plasma condenses into the cold disk, which will lead to a softer spectrum or a positive $\Gamma$-$L_{2-10keV}/L_{Edd}$ correlation. We did not find a possible correlation between the Fe K$\alpha$ line and the appearance/disappearance of BELs in these five CLAGNs. The origin of the narrow Fe K$\alpha$ line is unclear, which includes the outer region of accretion disk, the broadline region and the parsec-scale torus (e.g., Shu et al. 2012; LaMassa et al. 2017). No correlation between the Fe K$\alpha$ line and the BELs may suggest that their origin is different, where the Fe K$\alpha$ line may originate from a region far from the nucleus (e.g., the torus).

The BELs are widely observed in bright AGNs (e.g., Seyferts and QSOs), while LLAGNs only show weak H$\alpha$ BELs (Hermosa Mun˜oz et al. 2020). In CLAGNs, the broad lines show a strong evolution from type 1.0/1.2 to 1.8/1.9 or vice visa (e.g., Shappee et al. 2014; LaMassa et al. 2015). Our results show that CLAGNs with broad H$\beta$ lines always have $L_{bol}/L_{Edd} \geq 10^{-2}$; this result should be similar for traditional bright AGNs (Seyferts and QSOs). The broad H$\beta$ lines are quickly become weak or disappear when $L_{bol}/L_{Edd} \leq 10^{-2}$, which is quite similar to the behavior in LLAGNs. The origin and evolution of BELs is still not so clear in AGNs (e.g., Czerny & Hryniewicz 2011; Elvis 2017; Czerny 2019). The CLAGNs provide a chance to explore the evolution of BELs. Our results on the coevolution of BELs and the X-ray spectrum in CLAGNs suggest that the evolution of the broad lines is probably regulated by the accretion process, which includes the variation of the ionization or the disk winds that possibly form the cloud of BELs. As the accretion rate decreases to lower than a critical value, the inner cold disk may transit to the hot ADAF. The decrease in high-energy UV/X-ray photons will lead to a weakness of the high-ionization lines (e.g., H$\beta$ lines), and the H$\alpha$ lines will also disappear if the transition radius of the disk increases further. It should be noted that the dynamics of the BLR will strongly evolve with the possible disk transition because the radiation pressure force on the cloud will decrease or increase quickly when the transition radius of accretion disk increases or decreases. Therefore, we expect that the line profiles in CLAGNs will show strong variations during the changing-look process. Several CLAGNs also show a strong evolution of the BEL profiles, and the asymmetric or even double-peaked broad lines can appear. A change in the broad H$\beta$ line profile was found in NGC 3516 (Shapovalova et al. 2019). Oknyansky et al. (2021) suggest that inflowing BLR clouds exist based on the study of the double-peaked H$\beta$ lines with the reverberation mapping method. In the future, monitoring the broad lines during the changing-look phase of AGNs will further shed light on the physics of broad lines in AGNs.

## ACKNOWLEDGEMENTS

We thank the anonymous referee for their constructive suggestions that helped to the improve the manuscript. This work is supported by the NSFC (grants U1931203).

*Facility:* NuSTAR(FPMA and FPMB), XMM-Newton(PN, MOS1 and MOS2), Chandra(ACIS-S).
*Software:* HEASOFT (v6.28), XSPEC (v12.11.1), SAS (v19.1.0), CIAO (v4.11).

Table 2. The Fitting Results of the CLAGN Samples

| ObsID | Tel | UT Date | Type | References | Model | POW/PEXRAV Γ | POW/PEXRAV Norm($10^{-3}$) | ZGAUSS LE(keV) | ZGAUSS LW(keV) | ZGAUSS Norm($10^{-5}$) | $\chi^2$/dof. | log($L_X$/$L_{Edd}$) |
|---|---|---|---|---|---|---|---|---|---|---|---|---|
| (1) | (2) | (3) | (4) | (5) | (6) | (7) | (8) | (9) | (10) | (11) | (12) | (13) |
| Mrk 1018 | | | | | | | | | | | | |
| 0201090201 | X | 2005-01-15 | BEL | M16 | a | $1.82^{+0.03}_{-0.04}$ | $2.59^{+0.11}_{-0.11}$ | - | - | - | 709.31/691 | -2.37 |
| 0554920301 | X | 2008-08-07 | BEL | M16 | a | $1.93^{+0.02}_{-0.01}$ | $3.84^{+0.06}_{-0.06}$ | - | - | - | 1291.21/1131 | -2.28 |
| 12868 | Ch | 2010-11-27 | BEL | K18 | a | $1.57^{+0.04}_{-0.04}$ | $0.67^{+0.02}_{-0.02}$ | - | - | - | 177.53/155 | -2.79 |
| 60160087002 | N | 2016-02-10 | NEL | L21 | b | $1.84^{+0.10}_{-0.10}$ | $0.52^{+0.10}_{-0.10}$ | - | - | - | 277.80/294 | -3.10 |
| 18789 | Ch | 2016-02-25 | NEL | L21 | a | $1.69^{+0.06}_{-0.06}$ | $0.27^{+0.02}_{-0.01}$ | - | - | - | 174.09/181 | -3.26 |
| 19560 | Ch | 2017-02-17 | NEL | L21 | a | $1.64^{+0.04}_{-0.03}$ | $0.49^{+0.02}_{-0.01}$ | - | - | - | 379.97/314 | -2.98 |
| 60301022002 | N | 2018-01-05 | NEL | L21 | b | $1.66^{+0.07}_{-0.07}$ | $0.51^{+0.08}_{-0.07}$ | - | - | - | 359.88/388 | -2.99 |
| 20366 | Ch | 2018-01-08 | NEL | L21 | a | $1.57^{+0.07}_{-0.07}$ | $0.33^{+0.02}_{-0.02}$ | - | - | - | 177.11/171 | -3.10 |
| 20367 | Ch | 2018-03-03 | NEL | L21 | a | $1.69^{+0.07}_{-0.07}$ | $0.33^{+0.02}_{-0.02}$ | - | - | - | 136.54/153 | -3.19 |
| 60301022003 | N | 2018-03-05 | NEL | L21 | b | $1.80^{+0.08}_{-0.09}$ | $0.41^{+0.07}_{-0.07}$ | - | - | - | 510.63/514 | -3.18 |
| 20368 | Ch | 2018-06-12 | NEL | L21 | a | $1.66^{+0.07}_{-0.07}$ | $0.36^{+0.02}_{-0.02}$ | - | - | - | 178.63/169 | -3.13 |
| 60301022005 | N | 2018-07-17 | NEL | L21 | b | $1.61^{+0.06}_{-0.07}$ | $0.40^{+0.06}_{-0.05}$ | - | - | - | 345.25/327 | -3.05 |
| 0821240201 | X | 2018-07-23 | NEL | L21 | a | $1.68^{+0.02}_{-0.02}$ | $0.46^{+0.01}_{-0.01}$ | - | - | - | 910.38/884 | -3.03 |
| 20369 | Ch | 2018-09-09 | NEL | L21 | a | $1.60^{+0.06}_{-0.05}$ | $0.51^{+0.03}_{-0.02}$ | - | - | - | 227.85/216 | -2.93 |
| 20370 | Ch | 2018-11-08 | NEL | H20 | a | $1.66^{+0.08}_{-0.07}$ | $0.28^{+0.02}_{-0.02}$ | - | - | - | 142.86/147 | -3.23 |
| 0821240301 | X | 2019-01-04 | NEL | H20 | a | $1.65^{+0.03}_{-0.02}$ | $0.21^{+0.01}_{-0.01}$ | - | - | - | 450.12/433 | -3.35 |
| 21432 | Ch | 2019-02-06 | NEL | H20 | a | $1.67^{+0.06}_{-0.06}$ | $0.36^{+0.02}_{-0.02}$ | - | - | - | 209.22/186 | -3.14 |
| 22082 | Ch | 2019-02-07 | NEL | H20 | a | $1.65^{+0.07}_{-0.06}$ | $0.38^{+0.02}_{-0.02}$ | - | - | - | 184.50/176 | -3.10 |
| 21433 | Ch | 2019-10-09 | NEL | H20 | a | $1.70^{+0.06}_{-0.06}$ | $0.20^{+0.01}_{-0.01}$ | - | - | - | 186.65/185 | -3.40 |
| NGC 3516† | | | | | | | | | | | | |
| 0107460601 | X | 2001-04-10 | BEL | S19 | c | $1.72^{+0.02}_{-0.02}$ | $6.82^{+0.22}_{-0.22}$ | $6.41^{+0.01}_{-0.01}$ | $0.06^{+0.02}_{-0.01}$ | $4.31^{+0.41}_{-0.42}$ | 1906.14/1942 | -2.90 |
| 0107460701 | X | 2001-11-09 | BEL | S19 | c | $1.44^{+0.02}_{-0.02}$ | $2.66^{+0.09}_{-0.08}$ | $6.41^{+0.01}_{-0.01}$ | $0.09^{+0.01}_{-0.01}$ | $5.75^{+0.32}_{-0.31}$ | 2556.64/2240 | -3.10 |
| 0401210401 | X | 2006-10-06 | BEL | S19 | c | $2.13^{+0.01}_{-0.01}$ | $29.23^{+0.79}_{-0.79}$ | $6.37^{+0.02}_{-0.02}$ | $0.09^{+0.02}_{-0.03}$ | $5.61^{+0.71}_{-0.84}$ | 2492.09/2375 | -2.53 |
| 0401210501 | X | 2006-10-08 | BEL | S19 | c | $2.08^{+0.02}_{-0.02}$ | $23.83^{+0.62}_{-0.60}$ | $6.37^{+0.02}_{-0.02}$ | $0.09^{+0.02}_{-0.02}$ | $6.09^{+0.63}_{-0.66}$ | 2774.93/2507 | -2.59 |
| 0401210601 | X | 2006-10-10 | BEL | S19 | c | $1.93^{+0.02}_{-0.01}$ | $15.32^{+0.43}_{-0.41}$ | $6.38^{+0.02}_{-0.01}$ | $0.09^{+0.02}_{-0.02}$ | $6.70^{+0.59}_{-0.60}$ | 2582.61/2369 | -2.69 |
| 0401211001 | X | 2006-10-12 | BEL | S19 | c | $2.07^{+0.02}_{-0.02}$ | $23.12^{+0.63}_{-0.61}$ | $6.40^{+0.02}_{-0.01}$ | $0.08^{+0.02}_{-0.02}$ | $5.71^{+0.63}_{-0.63}$ | 1235.78/1151 | -2.60 |
| 60002042002 | N | 2014-06-24 | NEL | S19 | d | $1.69^{+0.17}_{-0.10}$ | $1.07^{+0.17}_{-0.14}$ | $6.38^{+0.05}_{-0.05}$ | $0.15^{+0.09}_{-0.09}$ | $2.04^{+0.39}_{-0.38}$ | 426.62/426 | -3.65 |
| 60002042004 | N | 2014-07-11 | NEL | S19 | d | $1.83^{+0.21}_{-0.07}$ | $1.93^{+0.21}_{-0.18}$ | $6.36^{+0.03}_{-0.04}$ | $0.05^{+0.07}_{-0.05}$ | $1.61^{+0.30}_{-0.27}$ | 748.59/671 | -3.49 |
| 60302016002 | N | 2017-12-05 | NEL | S19 | d | $1.70^{+0.15}_{-0.09}$ | $1.38^{+0.31}_{-0.18}$ | $6.34^{+0.13}_{-0.47}$ | $0.01^{+0.13}_{-0.01}$ | $1.99^{+0.43}_{-0.37}$ | 266.98/320 | -3.55 |
| 60302016004 | N | 2017-12-07 | NEL | S19 | d | $1.69^{+0.11}_{-0.09}$ | $1.31^{+0.23}_{-0.18}$ | $6.35^{+0.08}_{-0.07}$ | $0.24^{+0.09}_{-0.09}$ | $2.35^{+0.53}_{-0.51}$ | 352.23/342 | -3.57 |
| 60302016006 | N | 2017-12-11 | NEL | S19 | d | $1.88^{+0.21}_{-0.09}$ | $1.63^{+0.46}_{-0.20}$ | $6.33^{+0.04}_{-0.04}$ | $0.01^{+0.09}_{-0.01}$ | $2.20^{+0.36}_{-0.39}$ | 302.35/330 | -3.57 |
| 60302016008 | N | 2017-12-26 | NEL | S19 | d | $1.81^{+0.10}_{-0.08}$ | $3.19^{+0.47}_{-0.31}$ | $6.31^{+0.08}_{-0.09}$ | $0.16^{+0.13}_{-0.16}$ | $2.37^{+0.82}_{-0.74}$ | 404.85/397 | -3.27 |
| 60302016010 | N | 2017-12-28 | NEL | S19 | d | $1.79^{+0.07}_{-0.06}$ | $3.22^{+0.35}_{-0.30}$ | $6.31^{+0.12}_{-0.09}$ | $0.07^{+0.13}_{-0.07}$ | $1.45^{+0.56}_{-0.48}$ | 548.83/568 | -3.26 |

Table 2 *continued*

Table 2 *(continued)*

| ObsID | Tel | UT Date | Type | References | Model | \multicolumn{2}{c}{POW/PEXRAV} | \multicolumn{3}{c}{ZGAUSS} | $\chi^2$/dof. | $\log(L_X/L_{Edd})$ |
| --- | --- | --- | --- | --- | --- | --- | --- | --- | --- | --- | --- | --- |
| | | | | | | $\Gamma$ | Norm($10^{-3}$) | LE(keV) | LW(keV) | Norm($10^{-5}$) | | |
| (1) | (2) | (3) | (4) | (5) | (6) | (7) | (8) | (9) | (10) | (11) | (12) | (13) |
| 60302016012 | N | 2017-12-30 | NEL | S19 | d | $1.84^{+0.07}_{-0.06}$ | $4.42^{+0.25}_{-0.39}$ | $6.33^{+0.05}_{-0.08}$ | $0.05^{+0.06}_{-0.05}$ | $1.61^{+0.48}_{-0.48}$ | 616.31/622 | -3.15 |
| 08545911101/60160001002 | X/N | 2020-04-20 | BEL | O21 | d | $1.96^{+0.03}_{-0.03}$ | $12.72^{+0.66}_{-0.40}$ | $6.36^{+0.01}_{-0.02}$ | $0.03^{+0.03}_{-0.03}$ | $3.84^{+0.53}_{-0.58}$ | 2142.46/1937 | -2.76 |
| \multicolumn{13}{c}{Mrk 590} |
| 0109130301 | X | 2002-01-01 | BEL | D14 | a | $1.74^{+0.04}_{-0.04}$ | $1.13^{+0.04}_{-0.04}$ | | | | 636.29/661 | -2.75 |
| 4924 | Ch | 2004-07-03 | BEL | D14 | c | $1.65^{+0.05}_{-0.04}$ | $1.13^{+0.05}_{-0.05}$ | $6.99^{+0.34}_{-0.20}$ | $0.78^{+0.36}_{-0.21}$ | $6.18^{+2.65}_{-1.63}$ | 199.84/157 | -2.64 |
| 0201020201 | X | 2004-07-04 | BEL | D14 | c | $1.73^{+0.01}_{-0.01}$ | $1.66^{+0.02}_{-0.02}$ | $6.40^{+0.04}_{-0.02}$ | $0.05^{+0.04}_{-0.05}$ | $1.21^{+0.25}_{-0.24}$ | 1335.65/1308 | -2.57 |
| 15647 | Ch | 2013-06-16 | NEL | D14 | c | $1.83^{+0.18}_{-0.17}$ | $0.20^{+0.03}_{-0.04}$ | $6.42^{+0.13}_{-0.06}$ | $0.08^{+0.14}_{-0.08}$ | $0.75^{+0.35}_{-0.35}$ | 72.07/86 | -3.53 |
| 16109 | Ch | 2014-11-15 | NEL | D14 | c | $1.75^{+0.11}_{-0.11}$ | $0.24^{+0.03}_{-0.04}$ | $6.65^{+0.17}_{-0.20}$ | $0.36^{+0.19}_{-0.17}$ | $1.26^{+0.49}_{-0.45}$ | 202.95/214 | -3.38 |
| 60160095002 | N | 2016-02-05 | U | - | b | $1.70^{+0.27}_{-0.17}$ | $0.70^{+0.29}_{-0.16}$ | | | | 370.42/408 | -2.93 |
| 90201043002 | N | 2016-12-02 | U | - | b | $1.75^{+0.12}_{-0.11}$ | $0.81^{+0.15}_{-0.12}$ | | | | 806.53/762 | -2.90 |
| 80402610002 | N | 2018-10-27 | U | - | b | $1.64^{+0.05}_{-0.04}$ | $2.08^{+0.19}_{-0.03}$ | | | | 691.97/695 | -2.43 |
| \multicolumn{13}{c}{NGC 2992} |
| 0147920301‡ | X | 2003-05-19 | U | - | c | $1.74^{+0.01}_{-0.02}$ | $26.22^{+0.56}_{-0.51}$ | $6.40^{+0.05}_{-0.07}$ | $0.40^{+0.14}_{-0.09}$ | $25.73^{+6.58}_{-4.49}$ | 3115.99/3044 | -2.66 |
| 0654910301 | X | 2010-05-06 | U | - | c | $1.62^{+0.01}_{-0.02}$ | $1.30^{+0.02}_{-0.03}$ | $6.39^{+0.02}_{-0.01}$ | $0.08^{+0.02}_{-0.03}$ | $2.32^{+0.28}_{-0.28}$ | 2466.46/2385 | -3.88 |
| 0654910401 | X | 2010-05-16 | U | - | c | $1.61^{+0.02}_{-0.01}$ | $1.55^{+0.03}_{-0.03}$ | $6.40^{+0.01}_{-0.01}$ | $0.03^{+0.03}_{-0.03}$ | $2.11^{+0.23}_{-0.23}$ | 2552.58/2398 | -3.81 |
| 0654910501 | X | 2010-05-26 | U | - | c | $1.60^{+0.01}_{-0.01}$ | $2.83^{+0.04}_{-0.04}$ | $6.40^{+0.02}_{-0.02}$ | $0.01^{+0.01}_{-0.01}$ | $2.43^{+0.32}_{-0.24}$ | 3127.68/2970 | -3.54 |
| 0654910601 | X | 2010-06-05 | U | - | c | $1.62^{+0.02}_{-0.02}$ | $1.06^{+0.03}_{-0.03}$ | $6.41^{+0.01}_{-0.01}$ | $0.06^{+0.02}_{-0.02}$ | $2.55^{+0.25}_{-0.25}$ | 1390.77/1045 | -3.97 |
| 0654910701 | X | 2010-11-08 | U | - | c | $1.63^{+0.01}_{-0.02}$ | $1.12^{+0.02}_{-0.03}$ | $6.41^{+0.01}_{-0.01}$ | $0.05^{+0.03}_{-0.03}$ | $2.61^{+0.24}_{-0.25}$ | 1479.05/1112 | -3.95 |
| 0654910801 | X | 2010-11-18 | U | - | c | $1.66^{+0.02}_{-0.02}$ | $0.94^{+0.03}_{-0.02}$ | $6.40^{+0.01}_{-0.01}$ | $0.05^{+0.01}_{-0.03}$ | $2.71^{+0.23}_{-0.24}$ | 1921.33/1556 | -4.04 |
| 0654910901 | X | 2010-11-28 | U | - | c | $1.64^{+0.03}_{-0.03}$ | $0.55^{+0.02}_{-0.02}$ | $6.39^{+0.01}_{-0.01}$ | $0.07^{+0.02}_{-0.02}$ | $2.25^{+0.21}_{-0.22}$ | 1566.63/984 | -4.25 |
| 0654911001 | X | 2010-12-08 | U | - | c | $1.60^{+0.01}_{-0.02}$ | $1.20^{+0.02}_{-0.02}$ | $6.41^{+0.01}_{-0.01}$ | $0.06^{+0.03}_{-0.03}$ | $2.20^{+0.26}_{-0.25}$ | 1353.88/1186 | -3.90 |
| 0701780101 | X | 2013-05-11 | BEL | G21 | c | $1.59^{+0.02}_{-0.02}$ | $3.41^{+0.09}_{-0.08}$ | $6.41^{+0.02}_{-0.03}$ | $0.08^{+0.04}_{-0.04}$ | $4.39^{+0.87}_{-0.83}$ | 821.40/848 | -3.45 |
| 60160371002 | N | 2015-12-02 | BEL | G21 | d | $1.72^{+0.03}_{-0.03}$ | $15.96^{+0.76}_{-0.71}$ | $6.27^{+0.09}_{-0.10}$ | $0.25^{+0.13}_{-0.12}$ | $8.16^{+2.30}_{-2.10}$ | 948.83/1023 | -2.88 |
| 0840920201 | X | 2019-05-07 | BEL | G21 | c | $1.65^{+0.01}_{-0.00}$ | $21.65^{+0.09}_{-0.10}$ | $6.40^{+0.01}_{-0.01}$ | $0.08^{+0.02}_{-0.02}$ | $11.01^{+0.95}_{-0.87}$ | 5370.64/4297 | -2.70 |
| 08409 20301/90501623002 | X/N | 2019-05-10 | BEL | G21 | d | $1.60^{+0.01}_{-0.01}$ | $16.82^{+0.02}_{-0.07}$ | $6.40^{+0.01}_{-0.01}$ | $0.12^{+0.02}_{-0.03}$ | $11.34^{+1.00}_{-0.82}$ | 7118.68/5753 | -2.78 |
| \multicolumn{13}{c}{NGC 1566} |
| 0763500201 | X | 2015-11-05 | U | - | c | $1.75^{+0.01}_{-0.02}$ | $0.61^{+0.01}_{-0.01}$ | $6.41^{+0.03}_{-0.04}$ | $0.07^{+0.06}_{-0.07}$ | $0.45^{+0.12}_{-0.11}$ | 1040.00/1125 | -3.88 |
| 0800840201/80301601002 | X/N | 2018-06-26 | BEL | O20 | b | $1.90^{+0.01}_{-0.01}$ | $16.33^{+0.13}_{-0.12}$ | | | | 5774.72/4745 | -2.54 |
| 0820530401/80401601002 | X/N | 2018-10-04 | BEL | O20 | d | $1.75^{+0.01}_{-0.01}$ | $3.87^{+0.02}_{-0.03}$ | $6.39^{+0.01}_{-0.01}$ | $0.06^{+0.02}_{-0.02}$ | $2.37^{+0.20}_{-0.20}$ | 4007.36/3583 | -3.07 |
| 0840800401/80502006002 | X/N | 2019-06-05 | NEL | O20 | d | $1.71^{+0.01}_{-0.01}$ | $2.52^{+0.02}_{-0.02}$ | $6.40^{+0.02}_{-0.02}$ | $0.08^{+0.06}_{-0.07}$ | $1.28^{+0.23}_{-0.21}$ | 3016.55/2695 | -3.24 |
| 60501031002 | N | 2019-08-08 | NEL | O20 | d | $1.74^{+0.04}_{-0.03}$ | $4.87^{+0.27}_{-0.25}$ | $6.38^{+0.05}_{-0.05}$ | $0.13^{+0.10}_{-0.13}$ | $2.45^{+0.57}_{-0.52}$ | 921.56/894 | -2.97 |
| 0851980101 | X | 2019-08-11 | NEL | O20 | a | $1.72^{+0.02}_{-0.02}$ | $2.61^{+0.06}_{-0.06}$ | | | | 1142.95/1113 | -3.24 |
| 60501031004 | N | 2019-08-18 | NEL | O20 | d | $1.72^{+0.04}_{-0.04}$ | $2.57^{+0.17}_{-0.16}$ | $6.35^{+0.08}_{-0.08}$ | $0.27^{+0.13}_{-0.11}$ | $2.01^{+0.52}_{-0.45}$ | 797.36/818 | -3.23 |
| 60501031006 | N | 2019-08-21 | NEL | O20 | d | $1.69^{+0.03}_{-0.04}$ | $2.76^{+0.15}_{-0.15}$ | $6.38^{+0.07}_{-0.07}$ | $0.24^{+0.11}_{-0.10}$ | $1.95^{+0.46}_{-0.41}$ | 866.76/877 | -3.19 |

Table 2 *continued*

Table 2 *(continued)*

| ObsID | Tel | UT Date | Type | References | Model | POW/PEXRAV | | ZGAUSS | | | $\chi^2$/dof. | log($L_X$ /$L_{Edd}$) |
|---|---|---|---|---|---|---|---|---|---|---|---|---|
| | | | | | | $\Gamma$ | Norm($10^{-3}$) | LE(keV) | LW(keV) | Norm($10^{-5}$) | | |
| (1) | (2) | (3) | (4) | (5) | (6) | (7) | (8) | (9) | (10) | (11) | (12) | (13) |

NOTE—Column (1): Observation ID. Column (2): Telescope for observation (N: NuSTAR; Ch: Chandra; X: XMM-Newton). Column (3): Observing date. Columns (4) and (5): Types classified according to the optical H$\beta$ emission line width and references (BEL for broad emission line, NEL for narrow emission line, and U for uncertainty). Column (6): Fitting models (a: pha(pow), b: pha(pexrav), c: pha(pow+zgau), and d: pha(pexrav+zgau)). Columns (7) and (8): Photon index ($\Gamma$) and normalization value in units of $ph\ cm^{-2}\ s^{-1}$ of the power-law component. Columns (9), (10), and (11): Line Energy, line width and normalization value (units of $ph\ cm^{-2}\ s^{-1}$) of the redshift Gaussian component. Column (12): Ratio of the statistic $\chi^2$ and the degrees of freedom. Column (13): Eddington ratio of the X-ray luminosity, where $L_X$ is the absorption-corrected luminosity in the 2-10 keV band.

The references of Column (5) are M16: McElroy et al. (2016), K18: Kim et al. (2018), L21: Lyu et al. (2021), H20: Hutsemékers et al. (2020), S19: Shapovalova et al. (2019), O21: Oknyansky et al. (2021), D14: Denney et al. (2014), G21: Guolo et al. (2021), and O20: Oknyansky et al. (2020).

†: The energy band of the XMM-Newton data is 3-10 keV.
‡: The hydrogen column density $N_H$ was set as a free parameter.

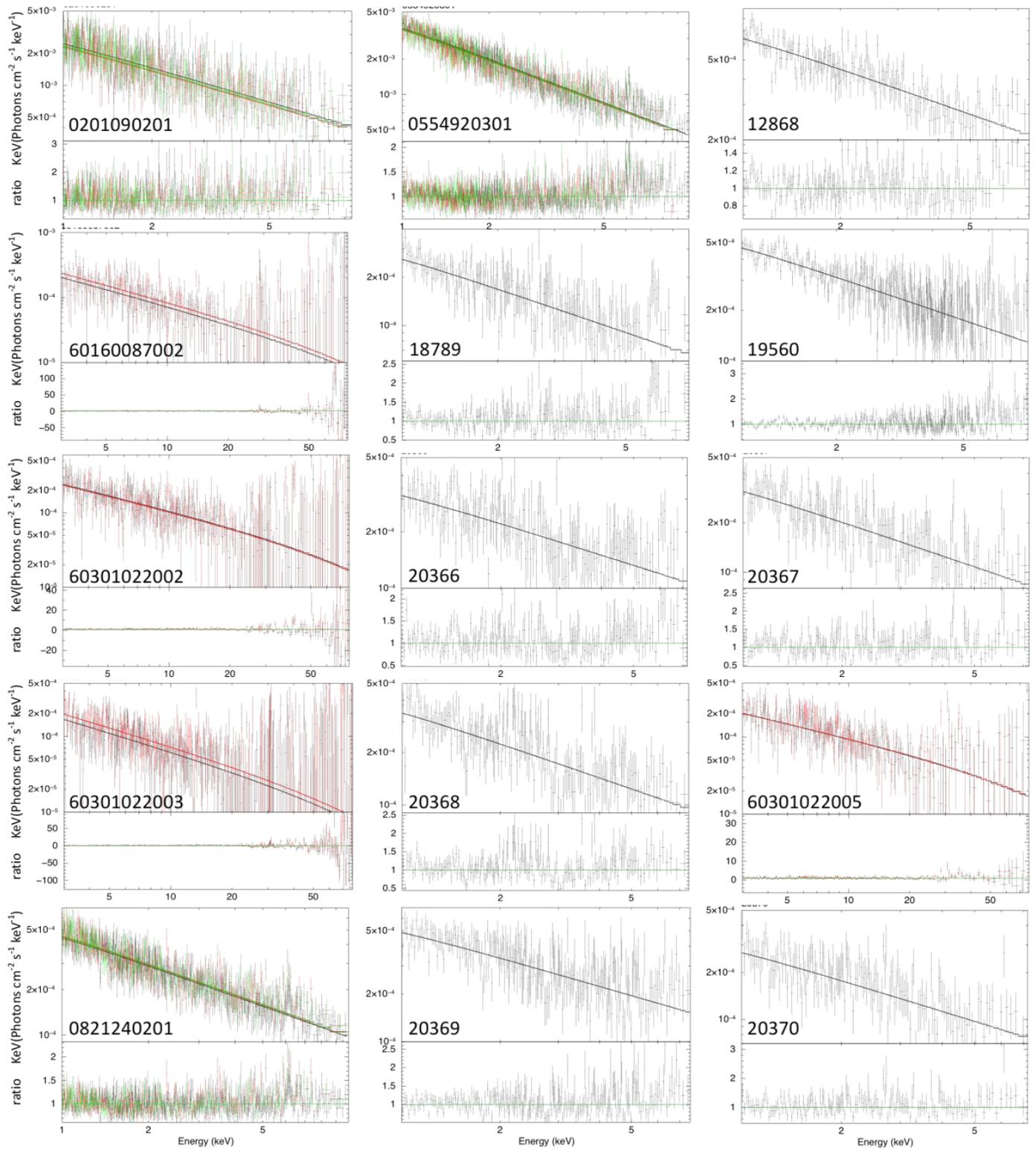

**Figure 3.** The unfolded X-ray spectra and their best-fitting models for 15 observations of Mrk 1018. The bottom part of each panel gives the ratio of data to model for each case.

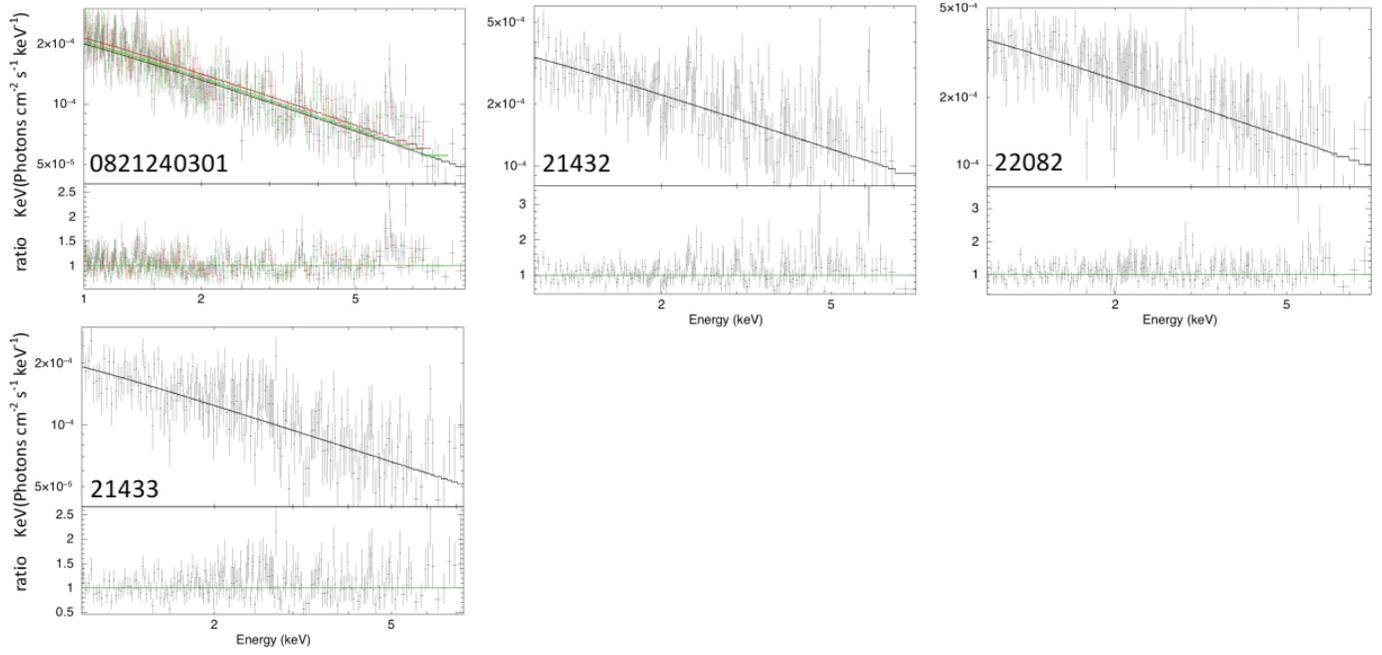

**Figure 4.** Same as Figure 3 but for another 4 observations of Mrk 1018.

16 Liu et al.

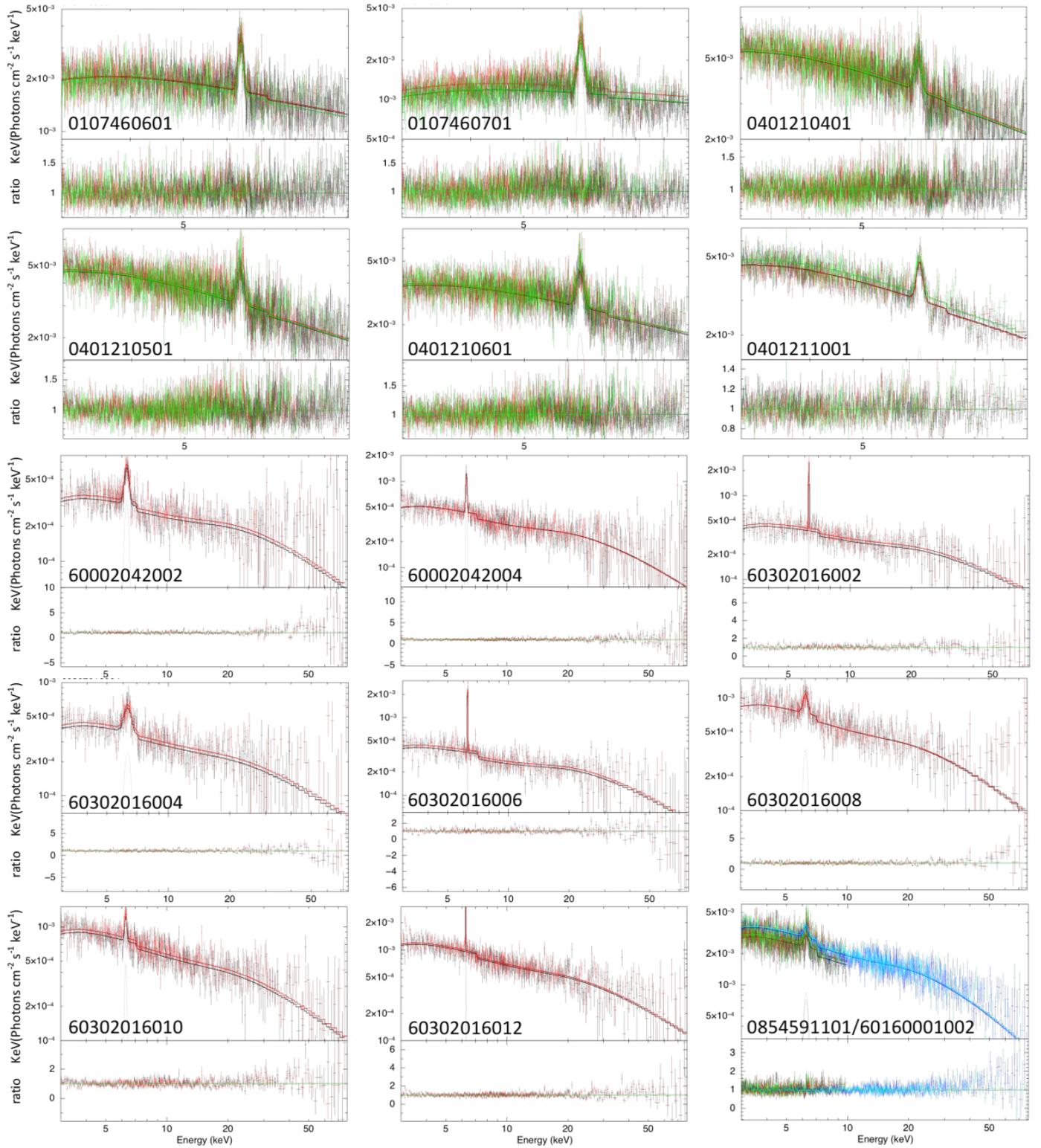

**Figure 5.** The unfolded X-ray spectra and their best fitting models for 15 observations of NGC 3516.

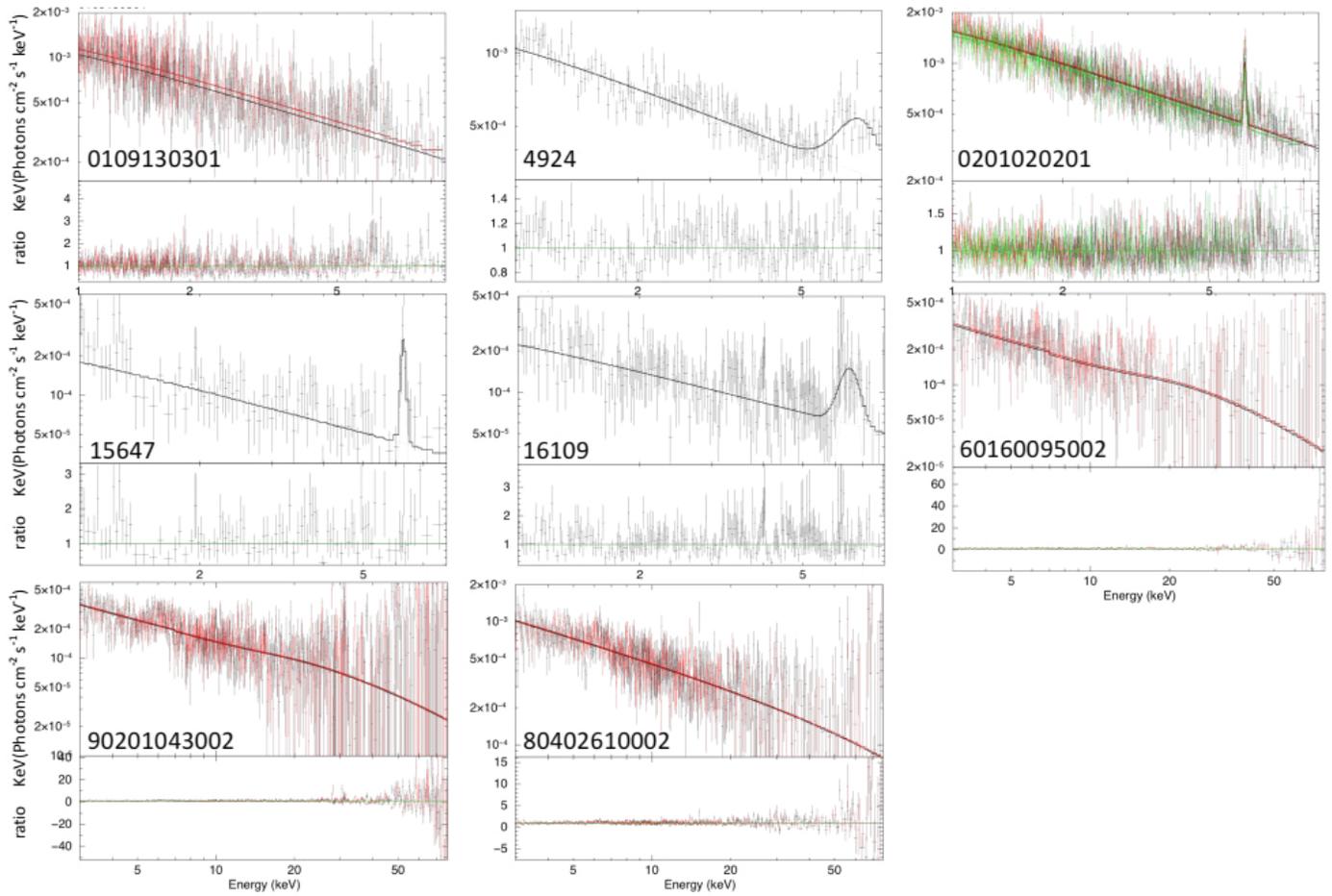

**Figure 6.** The unfolded X-ray spectra and their best fitting models for 8 observations of Mrk 590.

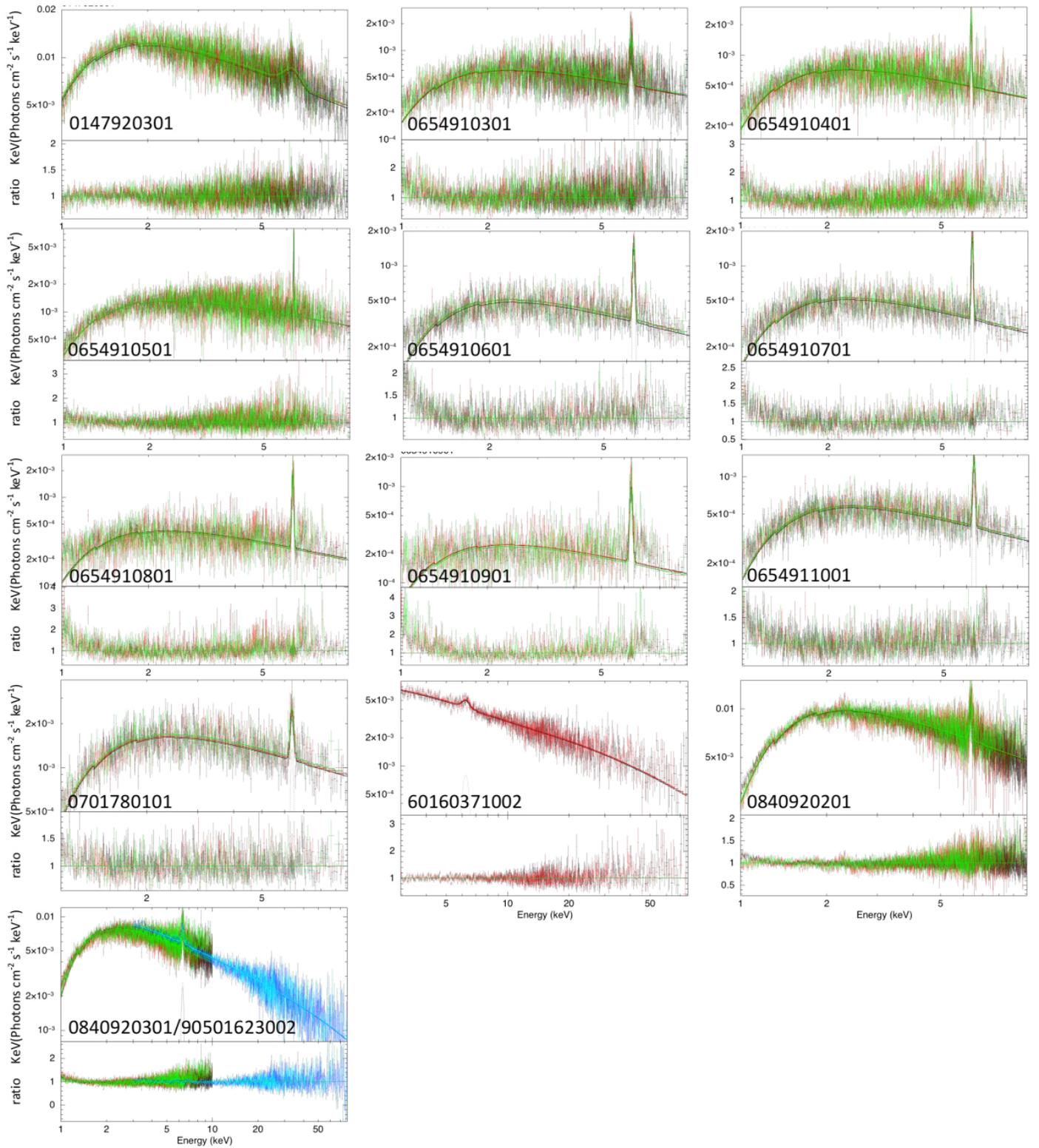

**Figure 7.** The unfolded X-ray spectra and their best fitting model for 13 observations of NGC 2992.

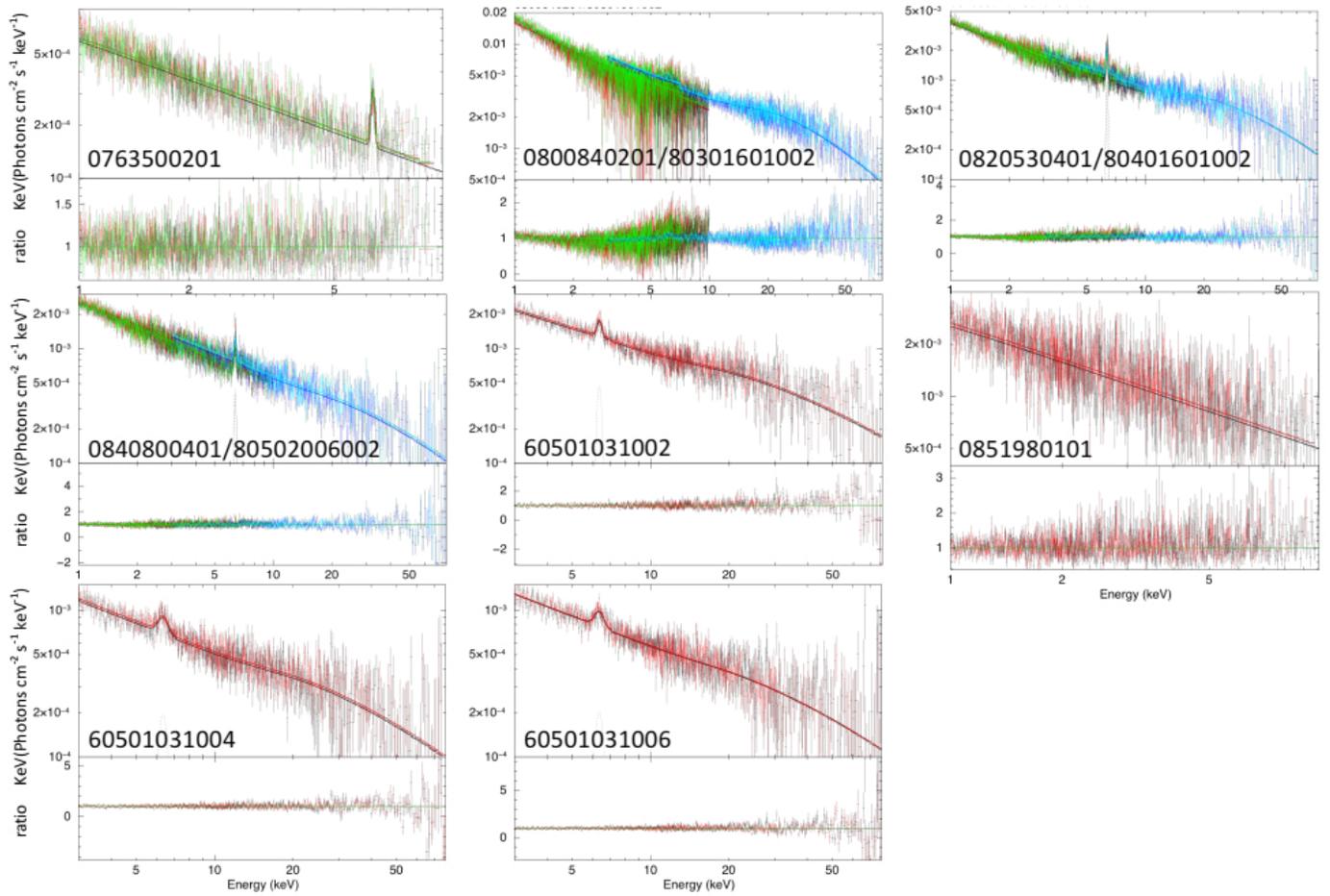

**Figure 8.** The unfolded X-ray spectra and their best fitting model for 8 observations of NGC 1566.